\documentstyle[pra,aps]{revtex}
\begin{document}
\title{Local environment can enhance fidelity of quantum teleportation}
\author{Piotr Badzi{\c{a}}g$^{1,}$\cite{poczta}, Micha\l {} Horodecki$^{2,}$
\cite{poczta1}, Pawe\l{} Horodecki$^{3,}$\cite{poczta2} and Ryszard Horodecki $%
^{2,}$\cite{poczta3}}
\address{$^1$ Department of Mathematics and Physics,
Malardalens Hogskola, S-721 23 Vasteras, Sweden, \\
$^2$ Institute of Theoretical Physics and Astrophysics,
University of Gda\'nsk, 80--952 Gda\'nsk, Poland,\\
$^3$Faculty of Applied Physics and Mathematics,
Technical University of Gda\'nsk, 80--952 Gda\'nsk, Poland\\
}
\maketitle

\begin{abstract}
We show how an interaction with the environment can enhance fidelity of
quantum teleportation. To this end, we present examples of states which
cannot be made useful for teleportation by any local unitary
transformations; nevertheless, after being subjected to a dissipative
interaction with the local environment, the states allow for teleportation
with genuinely quantum fidelity. The surprising fact here is that the
necessary interaction does not require any intelligent action from the
parties sharing the states. In passing, we produce some general results
regarding optimization of teleportation fidelity by local action. We show
that bistochastic processes cannot improve fidelity of two-qubit states. We
also show that in order to have their fidelity improvable by a local
process, the bipartite states must violate the so-called reduction criterion
of separability.
\end{abstract}

\section{Introduction}

Quantum teleportation \cite{Bennett_tel} is fundamentally important as an
operational test of the presence and the strength of entanglement. Moreover,
a recent series of beautiful experiments \cite{exp}, which realized
teleportation in practice, opened a window for a wide range of its possible
technological applications.

In this paper, teleportation is understood as any strategy which uses local
quantum operations and classical communication (LOCC) \cite{Bennett_pur} to
transmit an unknown state via a shared pair. In an ideal teleportation
scheme, the EPR-channel is constituted by a pure, maximally entangled
bipartite state:
\begin{equation}
\psi _{-}={\frac{1}{\sqrt{2}}}(|01\rangle -|10\rangle ).
\end{equation}
The state is shared by a sender (Alice) and a receiver (Bob). By sharing $%
\psi _{-}$ with Alice, Bob can produce an {\em exact} replica of another
(input) state originally held by Alice. In reality, however, interactions
with the environment and imperfections of preparation result in Alice and
Bob sharing a state which is always mixed. Consequently, at Bob's end, the
teleported state can only be a distorted copy of the input initially held by
Alice. Moreover, if the bipartite state is mixed too much, it will not
provide for any better transmission fidelity than that of an ordinary
classical communication channel \cite{Popescu94}. To do better than a
classical channel, the shared quantum state must be entangled. A natural
question then is \cite{Popescu94}: can any entangled state provide better
than classical fidelity of teleportation?

Early attempts to answer this question, concentrated on the characterization
of the states which can offer non-classical fidelity within the original
teleportation scheme supplemented by local unitary rotations. Henceforth we
will call such a scheme the {\em standard teleportation scheme} (STS).
Fidelity of teleportation achievable in STS is uniquely determined by the
bipartite state's {\it fully entangled fraction}. It was defined in \cite
{huge} as
\begin{equation}
f(\varrho )=\max_{\psi }\langle \psi |\varrho |\psi \rangle .  \label{fully}
\end{equation}
In the definition, the maximum is taken over all maximally entangled states $%
\psi $ i.e. over $\psi =U_{1}\otimes U_{2}\psi _{+}$, where
\begin{equation}
\psi _{+}={\frac{1}{\sqrt{d}}}\sum_{i=1}^{d}|i\rangle |i\rangle  \label{plus}
\end{equation}
$U_{1}$ and $U_{2}$ are unitary transformations. Later, it was shown that in
order to be useful for STS, the states acting on a Hilbert space $%
C^{d}\otimes C^{d}$ must have $f>1/d$ \cite{tel,single}. Moreover, it was
shown that no {\it bound entangled} state (see \cite{bound}) can offer
better fidelity than classical communication \cite{Linden,single}. Somewhat
earlier, in Refs. \cite{Massar,Kent}, the authors identified a class of
states which do not permit any increase of $f$, neither by any trace
preserving (TP) LOCC nor even by some less restricted non-TP LOCC actions.
Mixtures of a maximally mixed state and $\psi _{+}$ \cite{Popescu94,Werner}
belong, among others, to this class.

One could then be tempted to speculate that $f$ could not be increased by
any TP LOCC operations. If so, then STS would be a unique
teleportation scheme in the sense that no other scheme would provide better
fidelity than STS. On the other hand, one could still suspect that by some
intelligent, sophisticated LOCC operation, Alice and Bob would be able to
increase $f$ for some states anyway. An important question was then to be
answered:

{\em Is it possible to design a teleportation scheme, for which at least
some states with }$f\leq 1/d${\em \ would give non-classical fidelity? }

In this paper, we answer this question by presenting a class of two-qubit
states with $f\leq 1/2$, which can, nevertheless, be used for teleportation
with non-classical fidelity. For that, however, one has to allow for some
{\em dissipative} interaction between the states and their local environment
first. This means that dissipation, which is usually associated with
decoherence and destruction of teleportation, increases $f$ of some
initially non-teleporting states to above $1/2$. In other words, some states
can produce non-classical fidelity within the original teleportation scheme
but only after being 'corrupted' by the environment !

To our knowledge, this is a previously unknown effect. In particular, it is
different than that used in the so called {\it filtering} method of
improving some of the states' parameters \cite{Gisin,conc}. Filtering
includes a selection process based on a {\it readout} of measurement
outcomes. In our examples, on the other hand, Alice and Bob do not need to
know the outcomes at all. Hence, in particular, unlike filtering, the actions in our
examples are entirely trace preserving.

We begin our presentation by recalling some of the general results on
optimal teleportation fidelity in Sect.\ref{sec2}(c.f. Ref. \cite{single}).
This allows us to conclude that an optimal teleportation scheme should
include maximization of $f$ by means of TP LOCC operations. Then, in Sect.
\ref{sec3} we put the problem in the context of increasing $f$ by the maps
of the form $I\otimes \Lambda $. We can limit the possible successful maps
by showing that, e.g., for two qubits, the bistochastic processes cannot do
the job. We also show that the states with $f$ improvable by $I\otimes
\Lambda $ action must violate the so called {\it reduction criterion}.
Subsequently, in Sect. \ref{sec4}\ we present the examples of states, for
which $f$ can be non-trivially increased by TP LOCC
operations. The paper ends with the summary of the results and the
conclusions in Sect. \ref{sec5}.

\section{Optimal fidelity in a general teleportation scheme}

\label{sec2} Let Alice and Bob share a pair of particles in a given state $%
\varrho $ acting on a Hilbert space ${\cal H}_{A}\otimes {\cal H}%
_{B}=C^{d}\otimes C^{d}$. Additionally, let Alice have a third particle in
an unknown pure state $\psi \in {\cal H}_{C}=C^{d}$ to be teleported. In the
most general teleportation scheme, Bob and Alice apply some trace preserving
(TP) (hence without selection of the ensemble) LOCC operation ${\cal T}$ to
the particles which they share and to the third (Alice's) particle. After
the operation is completed, the final state of Bob's particle (from the
pair) is
\begin{equation}
\varrho _{Bob}^{\psi }={\rm Tr}_{A,C}\left[ {\cal T}(|\psi \rangle \langle
\psi |\otimes \varrho )\right] .  \label{Bob}
\end{equation}
The resulting mapping of the input state (the state of the third particle)
onto $\varrho _{Bob}({\psi })$ establishes a {\it teleportation channel} $%
\Lambda $ (it depends on both, ${\cal T}$ and $\varrho $):
\begin{equation}
\Lambda (|\psi \rangle \langle \psi |)=\varrho _{Bob}(\psi ).
\end{equation}
The aim of teleportation is to bring $\varrho _{Bob}(\psi )$\ as close to $%
|\psi \rangle \langle \psi |$\ as possible. A useful measure of the quality
of teleportation is then provided by teleportation's {\em fidelity} \cite
{Popescu94}
\begin{equation}
{\cal F}=\overline{\langle \psi |\varrho _{Bob}(\psi )|\psi \rangle }.
\label{fidelity}
\end{equation}
Fidelity is a function of map $\Lambda $ and, like $\Lambda $,\ it depends
on both, teleporting state $\varrho $ and the strategy of teleportation $%
{\cal T}$ . One can show \cite{single} that in the standard teleportation
scheme, the maximal fidelity achievable from a given bipartite state $%
\varrho $ is
\begin{equation}
{\cal F}={\frac{fd+1}{d+1}}
\end{equation}
where $f$ is the fully entangled fraction of $\rho $\ given by formula (\ref
{fully}). To achieve this fidelity, Alice and Bob have to rotate their
respective parts of the teleporting state $\rho $ so that the maximum of
formula (2) is attained on singlet $\psi _{-}$. The original teleportation
scheme applied with the rotated bipartite state $\rho $\ will now produce
the maximal fidelity (8).

If, on the other hand, Alice and Bob do not share any quantum state, then
their best strategy is \cite{Popescu94}:

\begin{enumerate}
\item[(i)]  Alice performs an optimal measurement of the system to be
teleported and sends the outcome to Bob (classically).

\item[(ii)]  On the basis of her results, Bob tries to reconstruct the state.
\end{enumerate}

The optimal teleportation fidelity for this strategy is equal to the optimal
fidelity of the state estimation for a single system. It is given by \cite
{Massar95,single}
\begin{equation}
{\cal F}_{cl}={\frac{2}{1+d}}.
\end{equation}
One can easily see now that, in order to perform better than classical
communication, STS needs bipartite states with $f>1/d$. With $f\leq 1/d$,
Alice and Bob can just as well discard their bipartite state and communicate
classically.

There is no reason why STS should represent the most efficient teleportation
scheme using states with $f>1/d$. One can show, however, that the optimal
teleportation scheme (OTS) is a generalization of STS \cite{single}. OTS
consists of two steps:

\begin{enumerate}
\item[(i)]  Alice and Bob try to maximize $f$\ by applying TP
LOCC (not necessarily unitary) operations to the original state $\varrho $.

\item[(ii)]  They apply STS using the transformed state.
\end{enumerate}

Let then $f_{max}(\varrho )$ denote the maximal $f$ attainable from $\varrho
$ by means of TP LOCC operations. The maximal teleportation fidelity from
state $\varrho $ is then given by \cite{single}
\begin{equation}
{\cal F}_{max}={\frac{f_{max}d+1}{d+1}}.
\end{equation}
Thus, to find the optimal teleportation fidelity for a given bipartite state
$\rho $, one must find $f_{max}$. In other words, the fidelity of STS can be
improved if:

\begin{enumerate}
\item  $f$ can be increased by LOCC,

\item  The final $f$ is in quantum region i.e. it is greater than $1/d$.
\end{enumerate}

Henceforth, when referring to a process of increasing $f$, we will
understand it as increasing so that the final value is above $1/d$ (Within
the range $f\leq 1/d$, the fully entangled fraction can be increased
relatively easily. This, however, does not produce any better fidelity than $%
{\cal F}_{cl}$).

\section{\label{3}Some general results on improving ${\cal F}$ by local
intractions}

\label{sec3}

\subsection{A simplified formula for maximal $f$ attainable by local
interaction}

When local TP transformations are used to increase $f$ of a general
bipartite state $\varrho \in C^{d}\otimes C^{d}$, then the best attainable
result is
\begin{equation}
f_{A}=\max_{\Lambda }\mbox{Tr}\left( (\Lambda \otimes I)\varrho P_{+}\right)
.  \label{fa}
\end{equation}
The maximum is here taken over all TP completely positive (CP) maps $\Lambda
$ and $P_{+}=|\psi _{+}\rangle \langle \psi _{+}|$, with $\psi _{+}$ given
by (\ref{plus}). Stinespring decomposition of $\Lambda $\ gives \cite{Kraus}
\begin{equation}
\Lambda (\cdot )=\sum_{i}V_{i}(\cdot )V_{i}^{\dagger }
\end{equation}
with $\sum_{i}V_{i}^{\dagger }V_{i}=I$. Moreover, we can utilize the fact
that $A\otimes I\psi _{+}=I\otimes A^{T}\psi _{+}$ \cite{Jozsa} (superscript
$T$ denotes transposition in basis $\left\{ |i\rangle \right\} $) and
rewrite formula (\ref{fa}) as
\begin{equation}
f_{A}=\max_{\Gamma }\mbox{Tr}\left( \varrho (I\otimes \Gamma )P_{+}\right) ,
\end{equation}
with
\begin{equation}
\Gamma (\cdot )=\sum_{i}W_{i}(\cdot )W_{i}^{\dagger }
\end{equation}
and $W_{i}=V_{i}^{\ast }$ (the star denotes complex conjugation). Naturally,
like $\Lambda $, $\Gamma $ is trace preserving, too.

We can now recall that there is an isomorphism between the TP CP maps and
the bipartite states with one subsystem maximally mixed. The isomorphism is
given by
\begin{equation}
\varrho ^{\prime }=(I\otimes \Lambda )P_{+}.
\end{equation}
Thus, for any TP CP map, the corresponding state has a maximally mixed
subsystem $A$ and for any state with a maximally mixed subsystem $A$, there
exists a map that realizes it via the above formula. Consequently, we can
obtain the following form for $f_{A}$
\begin{equation}
f_{A}(\varrho )=\max_{\varrho ^{\prime }}\mbox{Tr}(\varrho \varrho ^{\prime
}),  \label{general}
\end{equation}
where the maximum is taken over all states $\varrho ^{\prime }$ with
maximally mixed subsystem $A$. An analogous formula holds for $f_{B}$. In
general, the values $f_{A}$ and $f_{B}$ are likely to be different from one
another.

Formula (\ref{general}) allows for identification of those maps which
definitely cannot improve $f$. Take, for instance, the maps describing the
action of random external fields \cite{Alicki}. They are of the form
\begin{equation}
\Lambda (\cdot )=\sum_{i}p_{i}U_{i}(\cdot )U_{i}^{\dagger },  \label{random}
\end{equation}
with $U_{i}$ denoting unitary transformations. The corresponding $\varrho
^{\prime }=(I\otimes \Lambda )P_{+}$ is a mixture of maximally entangled
vectors. Consequently, $\mbox{Tr}(\varrho \varrho ^{\prime })$ cannot exceed
$f(\varrho )$ which is equal to the maximal overlap of $\varrho $ with one
maximally entangled vector.

In addition to preserving trace, maps (\ref{random}) preserve the identity,
i.e. $\Lambda (I)=I$. Maps preserving both the trace and the identity are
called bistochastic. In general, the class of bistochastic maps can be wider
than the class specified by (\ref{random}). For two qubits, however, the two
classes coincide. To see this, one can note that, in general, the set of
states corresponding to the set of bistochastic maps via the isomorphism
consists of the states with {\it both} subsystems maximally mixed. For
two-qubit systems such states are mixtures of maximally entangled vectors
\cite{inf}. Each such vector can be written as $I\otimes U\psi _{+}$ for
some unitary $U$. Hence, the maps corresponding to mixtures of such vectors
are mixtures of unitary maps. Thus, for two qubits the bistochastic maps
cannot increase $f$. One may conjecture that this should be the case in
higher dimensions, too.

\subsection{Increasing $f$ by local actions and the reduction criterion for
separability}

Let us now derive some constraints for the states with $f$ improvable by
local interaction. A state suitable for a teleportation channel must be
entangled, i.e., it must be impossible to represent it by a mixture of
product states \cite{Werner}.
\begin{equation}
\varrho \neq \sum_{i}p_{i}\varrho _{i}\otimes \tilde{\varrho}_{i}.
\end{equation}
Such states violate different separability criteria. Here, we consider the
so called {\it reduction criterion} for separability. It is given by the
following conditions satisfied by all separable states \cite{cerf,xor}:
\begin{equation}
\varrho _{A}\otimes I-\varrho \geq 0,\quad I\otimes \varrho _{B}-\varrho
\geq 0.  \label{kryt}
\end{equation}
The inequalities mean that the operators on the left hand sides must be {\it %
positive}, i.e., they must have nonnegative eigenvalues only. In a two-qubit
case, the reduction criterion is equivalent to separability (hence it is
also a sufficient condition for separability), while it becomes a weaker
``detector'' of entanglement in higher dimensions. In other words, there
exist non-separable (entangled) states in higher dimensions which do not
violate the reduction criterion.

Suppose now that for some state $\varrho $ one has $f_{A}(\varrho
)>f(\varrho )$, i.e., $f$ can be improved by a local TP operation on
subsystem $A$. Naturally, we require that the improvement is non-trivial,
i.e., $f_{A}>1/d$. We will show now that this condition implies violation of
the reduction criterion. Indeed, since $f_{A}>1/d$, then there exists a
state $\varrho ^{\prime }$ whose one subsystem (say, $\varrho _{A}^{\prime }$%
) has maximal entropy and:
\begin{equation}
\mbox{Tr}(\varrho \varrho ^{\prime })>1/d.  \label{ineq}
\end{equation}
Maximum entropy means that $\varrho _{A}^{\prime }=I/d$. This implies $%
\mbox{Tr}((\varrho _{A}\otimes I)\varrho ^{\prime })=\mbox{Tr}\left( \varrho
_{A}\varrho _{A}^{\prime }\right) =1/d$. By putting this into inequality (%
\ref{ineq}), we obtain
\begin{equation}
\mbox{Tr}\left( (\varrho _{A}\otimes I-\varrho )\varrho ^{\prime }\right) <0
\end{equation}
The trace of a composition of two positive operators is nonnegative.
Operator $\varrho ^{\prime }$ is positive. Consequently, in order to satisfy
the last inequality, the operator $\varrho _{A}\otimes I-\varrho $ cannot be
positive.

Since all the entangled two-qubit states violate the reduction criterion,
the condition for improvability of $f$ derived above, does not put any new
restrictions on the class of states with improvable $f$ here \cite
{Massar,Kent}. Nevertheless, the condition should be useful while
investigating bipartite states in more dimensions. This is because not all
the entangled states there violate the reduction criterion.

%Finally note that the reduction criterion consists of two conditions
%involving different reductions. There can be an asymetry: a state may
%violate only one of these conditions.

\section{Beating the standard teleportation scheme}

\label{sec4}Before showing how to do better than STS, we will still need to
introduce some methods of dealing with the fully entangled fraction of
two-qubit states.

\subsection{Fully entangled fraction in the Hilbert-Schmidt representation}

An arbitrary state of a two-qubit system can be represented as
\begin{equation}
\varrho ={\frac{1}{4}}(I\otimes I+\bbox {r\cdot \sigma}\otimes I+I\otimes %
\bbox {s\cdot \sigma}+\sum_{m,n=1}^{3}t_{nm}\sigma _{n}\otimes \sigma _{m}).
\end{equation}
Here, $I$ stands for the identity operator, ${\bbox r}$ and ${\bbox s}$
belong to $R^{3}$, $\{\sigma _{n}\}_{n=1}^{3}$ are standard Pauli matrices, $%
\bbox{
r\cdot\sigma}=\sum_{i=1}^{3}r_{i}\sigma _{i}$. Coefficients $t_{mn}={\rm Tr}%
(\rho \sigma _{n}\otimes \sigma _{m})$ form a real $3\times 3$ matrix later
denoted by $T$. Note that $\bbox r$ and $\bbox s$ are local parameters as
they determine the reductions of~$\varrho $:
\begin{eqnarray}
\varrho _{1} &\equiv &{\rm Tr}_{{\cal H}_{2}}\varrho ={\frac{1}{2}}(I+\bbox{
r\cdot\sigma}),  \nonumber \\
\varrho _{2} &\equiv &{\rm Tr}_{{\cal H}_{1}}\varrho ={\frac{1}{2}}(I+%
\bbox{s\cdot\sigma}).  \label{redukcje}
\end{eqnarray}
Matrix $T$ , on the other hand, is responsible for the correlations
\begin{equation}
E(\bbox{a},\bbox{b})\equiv \text{Tr}(\varrho \bbox{a\cdot\sigma}\otimes %
\bbox{b\cdot\sigma})=(\bbox a,T\bbox b).
\end{equation}
One can notice now, that for any two-qubit state $\varrho $, one can find a
product unitary transformation $U_{1}\otimes U_{2}$ which will transform $%
\varrho $ to a form with {\it diagonal} $T$. This statement follows from the
fact that for any $2\times 2$\ unitary transformation $U$, there is a unique
$3\times 3$\ rotation $O$ such that \cite{thir2}
\begin{equation}
U\bbox{ \hat n\cdot\sigma}U^{\dagger }=(O\bbox{\hat n})\bbox{\cdot\sigma}.
\label{pawel}
\end{equation}
Now, if a state is subjected to a $U_{1}\otimes U_{2}$ transformation, the
parameters $\bbox r,\bbox s$ and $T$ are transformed into
\begin{eqnarray}
&&\bbox r^{\prime }=O_{1}\bbox r,  \nonumber \\
&&\bbox s^{\prime }=O_{2}\bbox s,  \nonumber \\
&&T^{\prime }=O_{1}TO_{2}^{\dagger }.
\end{eqnarray}
with $O_{i}$'s corresponding to $U_{i}$'s via formula (\ref{pawel}). Thus,
for every two-qubit state $\rho $, we can always find such $U_{1}$ and $%
U_{2} $ so that the corresponding rotations will diagonalize $T$ \cite
{correl}. Moreover, by selecting suitable rotations, one can make $t_{11}$
and $t_{22}$ non-positive. In what follows, the states with diagonal $T$ and
$t_{11},t_{22}\leq 0$ will be called {\em canonical.}

For the states with diagonal matrix $T$ (hence also for the canonical
states), the fully entangled fraction is given by (c.f.\cite{hab})
\begin{equation}
f=\left\{
\begin{array}{l}
{\frac{1}{4}}(1+{\sum_{i}|t_{ii}|})\quad {\rm \ if\ }\quad {\rm det}T\leq 0
\\*[1mm]
{\frac{1}{4}}\left( 1+{\max_{i\not=k\not=j}(|t_{ii}|+|t_{jj}|-|t_{kk}|)}%
\right) {\rm \ if\ }{\rm \ det}T>0
\end{array}
\right. .  \label{max}
\end{equation}
One can show now \cite{inf,hab} that if ${\rm det}T\geq 0$, then $f\leq 1/2$%
, i.e., $f$ belongs to the classical region. Thus, while analyzing $f$ in
the quantum region, it will be convenient to investigate a relatively simple
function $N(\varrho )$, instead of a more involved matrix $T$. Function $%
N(\varrho )$ is given by
\begin{equation}
N(\varrho )=\sum_{i}|t_{ii}|.
\end{equation}
It has the following important properties:

\begin{enumerate}
\item  $f(\varrho )={\frac{1}{4}}(1+N(\varrho ))$ for $f\geq {\frac{1}{2}}$

\item  $N(\varrho )\leq 1$ if and only if $f\leq {\frac{1}{2}}$
\end{enumerate}

It then contains all the information necessary to analyze $f$.

\subsection{The canonical form in terms of the matrix elements}

By applying the formula for $t_{ij}$, one can easily show that diagonality
of $T$ is equivalent to the following conditions for the matrix elements of $%
\varrho $ written in the standard basis ($|1\rangle =|00\rangle $, $%
|2\rangle =|01\rangle $ etc.):
\begin{eqnarray}
&&\varrho _{12}=\varrho _{34} \\
&&\varrho _{14}=\varrho _{32} \\
&&\varrho _{23}\text{\ and\ \ }\varrho _{14}\text{ are real}.
\end{eqnarray}
%The matrix elements of $\varrho $ are written here
Moreover, since $t_{11}=2(\varrho _{14}+\varrho _{23})$ and $%
t_{22}=2(\varrho _{23}-\varrho _{14})$, the condition $t_{11},t_{22}\leq 0$
is equivalent to
\begin{eqnarray}
&&\varrho _{23}\leq 0 \\
&&|\varrho _{23}|\geq \left| \varrho _{14}\right|
\end{eqnarray}
Thus, any state $\varrho $ can be locally rotated to a form with matrix
elements satisfying the above constraints. This gives the following
expression for $N(\varrho )$:
\begin{equation}
N(\varrho )=|1-2(\varrho _{22}+\varrho _{33})|-2\varrho _{23}.
\end{equation}
Now, for
\begin{equation}
\varrho _{22}+\varrho _{33}\geq {\frac{1}{2}}  \label{singlet-cond}
\end{equation}
we have $t_{33}\leq 0$ hence ${\rm det}T\leq 0$. Consequently, by eq. (\ref
{max}) the fully entangled fraction is given by
\begin{equation}
f(\varrho )={\frac{1}{4}}(1+N(\varrho ))={\frac{1}{2}}(\varrho _{22}+\varrho
_{33}-2\varrho _{23}).
\end{equation}
Then, with $-2\varrho _{23}$ large enough, one has $f\geq 1/2$ and $f$ \ is
attained on singlet $\psi _{-}$: $f=\langle \psi _{-}|\varrho |\psi
_{-}\rangle $.

\subsection{A local action which improves $f$.}

With the canonical form of $\varrho $\ at hand, it is not all that difficult
to eventually find examples of states with improvable $f$. After some
trials, we focused our attention on a simple family of states which in their
canonical form have $\varrho _{24}=\varrho _{13}=0$:
\begin{equation}
\varrho =\left[
\begin{array}{cccc}
\varrho _{11} & 0 & 0 & \varrho_{14} \\
0 & \varrho _{22} & -p_{23} & 0 \\
0 & -p_{23} & \varrho _{33} & 0 \\
\varrho_{14} & 0 & 0 & \varrho _{44}
\end{array}
\right]
\end{equation}
Here $p_{23}\geq 0$ and $\varrho_{14}$ is real. We assumed also that $%
\varrho $ satisfies the condition (\ref{singlet-cond}) and that $p_{23}\geq
(1-\varrho _{22}-\varrho _{33})/2$, so that the state has $f=\langle \psi
_{-}|\varrho |\psi _{-}\rangle \geq 1/2$. Explicitly, $f$ is given by
\begin{equation}
f(\varrho )={\frac{1}{2}}(\varrho _{22}+\varrho _{33}+2p_{23}).
\end{equation}

We know (see Sec.\ref{sec3}) that bistochastic maps cannot improve $f$. So,
to improve it, we must try a non-bistochastic map. A possible simple
candidate is, e.g., a map which acts on Bob's qubit and transforms it as
follows:
\begin{equation}
\varrho _{B}\rightarrow \tilde{\varrho}_{B}=\Lambda (\varrho )=W_{0}\varrho
_{B}W_{0}^{\dagger }+W_{1}\varrho _{B}W_{1}^{\dagger }  \label{action}
\end{equation}
where the operators $W_{i}$ are given by
\begin{equation}
W_{1}=\left[
\begin{array}{cc}
1 & 0 \\
0 & \sqrt{p}
\end{array}
\right] ,\quad W_{2}=\left[
\begin{array}{cc}
0 & \sqrt{1-p} \\
0 & 0
\end{array}
\right]
\end{equation}
It is easy to check that $W_{i}$'s satisfy $W_{1}^{\dagger
}W_{1}+W_{2}^{\dagger }W_{2}=I$, hence the operation is trace preserving.
Moreover, one can notice that $\Lambda $ can be regarded as resulting from
the interaction of a two-level atom (Bob's qubit) with electromagnetic field
(an environment). Such an interaction produces the following transitions:
\begin{equation}
|0\rangle _{a}|0\rangle _{e}\rightarrow |0\rangle _{a}|0\rangle _{e}
\end{equation}
\begin{equation}
|1\rangle _{a}|0\rangle _{e}\rightarrow \sqrt{p}|0\rangle _{a}|1\rangle _{e}+%
\sqrt{1-p}|1\rangle _{a}|0\rangle _{e}.
\end{equation}
where the subscripts $a$ and $e$ denote atomic and field states
respectively. The parameter $p$ is then interpreted as the probability of
photon emission from the atom in its upper state $|1\rangle _{a}$. This kind
of interaction is called the {\it amplitude damping channel} and one can
check \cite{Preskill} that, if repeatedly applied to a qubit, it produces an
exponential decay characteristic to spontaneous emission. The completely
positive map $\Lambda $ is then obtained from the amplitude damping channel
by tracing out the environment variables \cite{Kraus}.

Let us then put $\sqrt{p}=\sin \theta $ and apply transformation (\ref
{action}) to Bob's part of the total (2-qubit) system. The 2-qubit operator
corresponding to $W_{i}$\ is $A_{i}\equiv I\otimes W_{i}$ and, consequently,
we obtain
\begin{equation}
\varrho \rightarrow \varrho ^{\prime }=A_{1}\varrho A_{1}^{\dagger
}+A_{2}\varrho A_{2}^{\dagger }  \label{oper}
\end{equation}
with
\begin{equation}
A_{1}=\left[
\begin{array}{cccc}
1 & 0 & 0 & 0 \\
0 & \cos \theta & 0 & 0 \\
0 & 0 & 1 & 0 \\
0 & 0 & 0 & \cos \theta
\end{array}
\right]
\end{equation}
and
\begin{equation}
A_{2}=\left[
\begin{array}{cccc}
0 & \sin \theta & 0 & 0 \\
0 & 0 & 0 & 0 \\
0 & 0 & 0 & \sin \theta \\
0 & 0 & 0 & 0
\end{array}
\right] .
\end{equation}
Note that like the original state $\varrho $, the new state $\tilde{\varrho}$
is in its canonical form, too.
\begin{equation}
\ \kern-6mm\tilde{\varrho}=\left[
\begin{array}{cccc}
\varrho _{11}+\varrho _{22}\sin ^{2}\theta & 0 & 0 & \varrho _{14}\cos \theta
\\
0 & \varrho _{22}\cos ^{2}\theta & -p_{23}\cos ^{2}\theta & 0 \\
0 & -p_{23}\cos ^{2}\theta & \varrho _{33}+\varrho _{44}\sin ^{2}\theta & 0
\\
\varrho _{14}\cos \theta & 0 & 0 & \varrho _{44}\cos ^{2}\theta
\end{array}
\right]
\end{equation}
The change of $f$ associated with the transformation is now given by $\Delta
_{B}=\langle \psi _{-}|\tilde{\varrho}|\psi _{-}\rangle -f(\varrho )$. A
simple calculation shows that
\begin{equation}
\Delta _{B}=\left( 1-\cos \theta \right) \left[ \frac{1+\cos \theta }{2}%
\left( \varrho _{44}-\varrho _{22}\right) -p_{23}\right] .  \label{3a}
\end{equation}
Here, the index $B$ indicates that Bob's qubit has been transformed. One can
check that if one transforms Alice's qubit instead of Bob's then the
resulting $\Delta _{A}$ is given by
\begin{equation}
\Delta _{A}=\left( 1-\cos \theta \right) \left[ \frac{1+\cos \theta }{2}%
\left( \varrho _{44}-\varrho _{33}\right) -p_{23}\right] .  \label{3b}
\end{equation}
Finally, one can swap places of $1$ and $\cos \theta $ on the diagonal of
the first transformation matrix $A_{1}$ and adjust $A_{2}$ accordingly.
This, translated into changes of $f$, result in expressions like (\ref{3a})
and (\ref{3b}) but with $\varrho _{44}$ substituted by $\varrho _{11}$. In
other words, single qubit, trace preserving transformations like that
defined by (\ref{oper}) can improve fidelity of states in form (29) provided
that
\begin{equation}
\left[ \max \left( \varrho _{11},\varrho _{44}\right) -\min \left( \varrho
_{22},\varrho _{33}\right) \right] -p_{23}\geq 0.  \label{eq4}
\end{equation}
The maximal increase $\Delta =\max \{\Delta _{A},\Delta _{B}\}$ achievable
in this way is
\begin{equation}
\Delta =\frac{\left[ \max \left( \varrho _{11},\varrho _{44}\right) -\min
\left( \varrho _{22},\varrho _{33}\right) -p_{23}\right] ^{2}}{2\left[ \max
\left( \varrho _{11},\varrho _{44}\right) -\min \left( \varrho _{22},\varrho
_{33}\right) \right] }
\end{equation}
To obtain a more clear picture of the situation, let us write the diagonal
elements of $\varrho $ as:
\begin{equation}
\varrho _{11}=\frac{1-\varepsilon -\gamma }{4}\quad \varrho _{44}=\frac{%
1-\varepsilon +\gamma }{4}
\end{equation}
\begin{equation}
\quad \varrho _{22}=\frac{1+\varepsilon -\delta }{4}\quad \varrho _{33}=%
\frac{1+\varepsilon +\delta }{4}
\end{equation}
To satisfy $\left( \varrho _{22}+\varrho _{33}+2p_{23}\right) \geq 1$ (so
that $f(\varrho )=\langle \psi _{-}|\varrho |\psi _{-}\rangle \geq 1/2$),
one needs a non-negative $\varepsilon $ and :
\begin{equation}
\frac{1-\varepsilon }{4}\leq p_{23}\leq \frac{1}{4}\sqrt{\left(
1+\varepsilon \right) ^{2}-\delta ^{2}}.
\end{equation}
(the upper limit for $p_{23}$\ guaranties positivity of $\varrho $). Thus,
the method improves $f$ on states with $0<\varepsilon <1$ and $\left| \gamma
\right| +\left| \delta \right| -2\varepsilon >4\,p_{23}$. One can easily
check that in this class, the ''most improvable'' border state ($%
4\,p_{23}=1-\varepsilon $, i.e., $f=1/2$) is
\begin{equation}
\varrho ={\frac{1}{2}}\left[
\begin{array}{cccc}
0 & 0 & 0 & 0 \\
0 & 3-2\sqrt{2} & 1-\sqrt{2} & 0 \\
0 & 1-\sqrt{2} & 1 & 0 \\
0 & 0 & 0 & 2\sqrt{2}-2
\end{array}
\right]  \label{stan}
\end{equation}
Since $f(\varrho )=1/2$ then standard teleportation scheme using $\varrho $\
does not offer any better fidelity than classical. On the other hand, if we
transform $\varrho $ by transformation (\ref{oper}) with $\cos \theta =(%
\sqrt{2}-1)/(4\sqrt{2}-5)$ (this choice maximizes $\Delta $), then the new
state still satisfies the condition (\ref{singlet-cond}), and we obtain $f(%
\tilde{\varrho})\approx 0.53>1/2$. The new state can than be used for
teleportation with non-classical fidelity
\begin{equation}
{\cal F}\approx {\frac{2.06}{3}}>{\frac{2}{3}}
\end{equation}
In other words, the state $\varrho $ gets ``better'' when corrupted by
environment. The improvement is small, nevertheless it is significant. It
changes the character of the state: from non-teleporting to teleporting.

While analyzing this result, one may notice that the states with the fully
entangled fraction improvable by the map (\ref{oper}) form a rather
restricted class. In particular, this map cannot increase the entangled
fraction of states like
\[
{\varrho}=\frac{1}{2}|\psi _{-}\rangle \langle \psi _{-}|+\frac{1}{2}%
|00\rangle \langle 00|.
\]
It would then be very interesting to provide a complete characterization the
class of states which allow to improve fidelity by some local process, as
well as the class of local processes capable to improve fidelity for some
states. This task is, however, beyond the scope of this paper.

\section{Conclusions}

\label{sec5}We have examined the problem of optimal teleportation fidelity
with given bipartite quantum states. To this end, we investigated a
possibility of increasing the fully entangled fraction by means of trace
preserving LOCC operations and discovered a class of LOCC operations\ \
which non-trivially increase $f$ on some of the two-qubit states. To a
surprise, the successful operations do not represent any sophisticated
action of Alice or Bob. Instead, they result from a common (dissipative)
interaction between the teleporting state and the local environment. The
unexpected conclusion then is that a dissipative interaction, normally
associated with the destruction of quantum teleportation, can sometimes
facilitate it.

P.B. acknowledges stimulating discussions with Richard Bonner and Benjamin
Baumslag. M.H., P.H. and R.H. are supported by Polish Committee for
Scientific Research, contract No. 2 P03B 103 16. P.B. is partially
supported by Svenska Institutet, project ML2000.

\end{document}